\documentclass[aps,nofootinbib,prd,eqsecnum,showpacs,showkeys,preprintnumbers]{revtex4-1}
\usepackage[caption=false]{subfig}
\usepackage{graphicx}
\usepackage{amsmath}
\usepackage{amsfonts}
\usepackage{amssymb}
\usepackage{color}
\usepackage{bm}
\usepackage{mathrsfs}
\usepackage{epstopdf}
\usepackage{url}
\usepackage{footnote}
\usepackage{textcomp}
\usepackage{ulem}
\usepackage{esint}
\usepackage[unicode=true, pdfusetitle,
bookmarks=true,bookmarksnumbered=false,bookmarksopen=false,
 breaklinks=false,pdfborder={0 0 1},backref=false,colorlinks=false]{hyperref}
\usepackage{multirow}
\usepackage{pifont}

\begin{document}
\title{Emergence of running vacuum energy in $f(R,T)$ gravity : Observational constraints}

\author{Ahmed Errahmani$^{1,2}$}
\email{ahmederrahmani1@yahoo.fr}
\author{Mounia Magrach$^{1}$}
\email{magrachmounia@gmail.com}
\author{Safae Dahmani$^{1,2}$}
\email{dahmani.safae.1026@gmail.com}
\author{Amine Bouali$^{1,2,3}$}
\email{a1.bouali@ump.ac.ma}
\author{Taoufik Ouali$^{1,2}$}
\email{t.ouali@ump.ac.ma}
\affiliation{$^{1}$Laboratory of Physics of Matter and Radiation, University of Mohammed first, BP 717, Oujda, Morocco\\$^{2}$Astrophysical and Cosmological Center, Faculty of Sciences, University of Mohammed first, BP 717, Oujda, Morocco\\$^{3}$Higher School of Education and Training, Mohammed I University, BP 717, Oujda, Morocco}
\date{\today}
\begin{abstract}
In this work, we present a new analysis for $f(R,T)$ gravity by exploring the energy momentum tensor. We demonstrate that $f(R,T)$ gravity with the form $f(R,T)=R+2 \kappa^2 \lambda T-2\Lambda$ is equivalent to Running Vacuum Energy (RVE), which interacts with the components of the cosmic fluid, namely dark matter and radiation. Interestingly, the form of such interaction is  inferred from the non-conservation of the stress energy tensor in $f(R, T)$ gravity  rather than being introduced in a phenomenological manner.  Furthermore,  the parameters that distinguish RVE from $\Lambda$CDM are fixed once the parameter of $f(R,T)$ gravity, $\lambda$, is known. To illustrate our setup, we perform a Markov Chain Monte Carlo analysis of three interaction scenarios using a combination of different data. we find that the  parameters characterizing the RVE model are  very small as expected. These results give an accuracy to this equivalence between $f(R,T)$ gravity under consideration  and support the recent result obtained from a quantum field theory in curved space-time point of view  which could open a new relationship between $f(R,T)$ gravity and quantum field theory. Finally, the interaction of the running vacuum increases the value of the current value of the Hubble rate by  $3.5\%$ compared to the $\Lambda$CDM model, which may be a promising study for the Hubble tension.  
\end{abstract}
\maketitle
\section{Introduction}
\label{intro}
Our Universe is currently undergoing an accelerated expansion, as evidenced by various cosmological observations such as Type Ia supernovae (SNIa) \cite{Riess1998,Filippenko1998,Perlmutter1999} cosmic microwave background anisotropies \cite{Ade2016}, and large scale structure \cite{Eisenstein2005,Blake2012,Parkinson2012,Kazin2014,Alam2015,Beutler2016}.  To account for this accelerated expansion, scientists have proposed the existence of an exotic dark energy (DE) component in the context of general relativity. The prevailing explanation for this accelerated expansion is the $\Lambda$ CDM model, in which CDM refers to cold dark matter and dark energy is ascribed to the cosmological constant, $\Lambda$, with an equation of state equal to $-1$. Even it remains the most consistent model supported by experimental data, this model faces some issues such as the fine tunning  and the coincidence problems. These disabilities led to the development of various models including quintessence \cite{Carroll1998,Ladghami2024}, phantom \cite{Johri2004,Sami2004,Bouali2021,Bouali2019,Mhamdi2023,Dahmani2023a,Dahmani2023b,Bouhmadi2015}, K-essence \cite{Malquarti2003,Chiba2000}, Quintom  \cite{Cai2010,Guo2005}, holography \cite{Horava2000,Li2004,Wang2017,Bouhmadi2018,Belkacemi2012,Bargach2021,Bouhmadi2011,Belkacemi2020,Enkhili2024a}, and running vacuum energy \cite{Sola2017,George2019,Shapiro2004,Rezaei2019}. 

On the other hand, an important modification that has attracted significant attention in the effort to clarify this expansion is the concept of modified gravity \cite{Bargach2021,Clifton2012,Mhamdi2024,Bouali2023,Errahmani2006}. Within this framework, the Lagrangian that usually describes gravitational effects, conventionally proportional to the scalar curvature, $R$, is replaced by a function that depends on that curvature.   One of the simpler modified theories of gravity is the $f(R)$ theory \cite{Starobinsky2007,Sotiriou2010,Sotiriou2009,Felice2010}. It is being explored as a potential alternative to general relativity for explaining the accelerated expansion of the Universe. 
There are other alternatives to general relativity (GR) by means  of the scalar torsion, $\mathcal{T}$, known as teleparallel equivalent to GR and referred as $f(\mathcal{T})$ gravity \cite{Capozziello2011,Bamba2013} or by the non-metricity, $Q$, dubbed as symmetric teleparallel equivalent to GR referred as $f(Q)$ gravity \cite{Mhamdi2024,Lazkoz2019,Mandal2020,Enkhili2024b}. Recently, other modified gravity theories have emerged by introducing the trace of the stress energy tensor of matter, $T$, such as $f(R,T)$ gravity \cite{Harko2011,Harko2014,Moraes2017,Errahmani2024},  $f(Q, T)$ gravity \cite{Xu2019,Xu2020,Arora2020}  and $f(G, T)$ gravity \cite{Elizalde2010,Cruz2012} where G is the Gauss-Bonnet scalar.

Furthermore, in the context of general relativity, dynamical vacuum energy density emerges from the explicit computation of quantum effects within Quantum Field Theory in curved space-time \cite{Moreno2022}. In these studies,  vacuum energy density is depicted as a series expansion, involving powers law  of the Hubble rate and/or its derivatives with respect to the cosmic time.  This kind of dynamical vacuum energy density is known as running vacuum energy (RVE)  \cite{Sola2017,George2019,Shapiro2004,Rezaei2019},  and have become an active area of research in theoretical physics and cosmology in recent years. The RVE model can explain the early inflationary phase of the Universe called  RVM-inflation \cite{Sola2022}, as well as the observed late-time acceleration \cite{Sola2023}. In the context of the late acceleration, the RVE structure is defined by a combination of the Hubble rate, $H$, and its derivative $\dot{H}$ \cite{Sola2017,Gomez2015}. Many recent studies conducted by \cite{Moreno2022,Rezaei2022} have shown that the equation of state for the running vacuum energy is not exactly equal to -1. Instead,  it may have dynamical behaviors and influenced by various factors associated with cosmic expansion within the framework of quantum field theory, i.e. the equation of state for the running vacuum energy evolves as a function of the Hubble rate and its derivatives.

Even though the basic idea of RVE was inspired by semi-qualitative of the renormalization group arguments \cite{Shapiro2009},  through an action functional approach \cite{Sola2008}, and recently in the framework of the quantum fields theory in curved space-time, RVE sourced from modified gravity, appears not to be highlighted in the literature at our knowledge and may be an appealing way to an interplay between  quantum field theory and modified gravity. This approach could offer a promising avenue for exploring the relationship between these two concepts. The purpose of this article is to establish an equivalence between modified gravity represented by $f(R,T)=R+2 \kappa^2 \lambda T-2\Lambda$ and an interacting running vacuum energy with the cosmic fluid in the context of general relativity. In this analysis, we address the  Friedmann equations, assuming that  the budget of the Universe consists of cold dark matter, radiation and RVE. We discuss four cases of interactions between the running vacuum energy density and the cosmic fluid components. In the first case,  the fluid contains both CDM and radiation (called model $A$) where only radiation interacts with RVE while CDM remains conserved. The second case,  model $B$,  CDM interacts with RVE while radiation remains conserved. Finally, model $C$,  all components interact with each other.

Following that, we perform a Markov Chain Monte Carlo (MCMC) analysis \cite{Padilla2021}, using the $H(z)$ measurements \cite{Zhang2014,Jimenez2002}, Pantheon+ \cite{Scolnic2022}, Baryon Acoustic Oscillation (BAO) \cite{Alam2017} and Cosmic Microwave Background (CMB) \cite{Aghanim2020} datasets in order to extract cosmological parameters from the three models and to assess the confidence level of the equivalence between $f(R,T)$ gravity and the interacting RVE. 
Furthermore, we conduct a comparison between our models and the $\Lambda$CDM model using information criteria, including the Akaike and Bayesian methods. 
Finally, we analyze the dynamics of the deceleration parameter, and the equation of state parameter across all models.

The structure of this work is as follows: In Section \eqref{Chap2}, we give an overview of $f(R,T)$ gravity. In Section \eqref{Chap3}, we present  the  Friedmann equations, the equivalence between $f(R,T)$ gravity and the running vacuum and the non-continuity equations. 
Section \eqref{Chap4} presents the three models considered in this work. 
In  Section \eqref{Chap5}, we describe the data used in our analysis. 
 Section \eqref{Chap6}  presents the results and discussions. 
Finally, Section \eqref{Chap8} is reserved for conclusions. In this article, we use the natural system of units {i.e. $G=c=1$, and we take the Einstein's gravitational constant as $\kappa^2=8 \pi G $.}

\section{A brief review of  $f(R,T)$ gravity}
\label{Chap2}
The $f(R,T)$ modified gravity being examined are derived from the subsequent Einstein-Hilbert action \cite{Harko2011}
\begin{equation}
S = \frac{1}{2\kappa^2}\int_{}^{}f(R,T)\sqrt{-g}d^4x + \int_{}^{}L_m\sqrt{-g}d^4x.
\label{1}
\end{equation}
In this context, $f(R,T)$ represents an arbitrary function of the Ricci scalar, $R$, and the trace of the stress-energy tensor of matter, $T$, $L_m$ is the Lagrangian density of matter and $g$ represents the determinant of the metric tensor, $g_{\mu\nu}$. We define the stress-energy tensor of matter as
\begin{equation}
T_{\mu\nu}= -\frac{2}{\sqrt{-g}}\frac{\delta(\sqrt{-g}L_m)}{\delta g^{\mu\nu}}.
\label{tkh}
\end{equation}
By varying the modified Einstein-Hilbert action Eq. \eqref{1} with respect to the metric tensor $g_{\mu\nu}$, we can express the gravitational field equations of $f(R,T)$ gravity as follows \cite{Harko2011}
\begin{align}
& F(R,T) {R}_{\mu \nu}-\frac{1}{2} f({R}, {T}) g_{\mu \nu}+\left(g_{\mu \nu} \square-\nabla_\mu \nabla_\nu\right) F({R}, {T}) \nonumber 
\\
&\qquad=\left(\kappa^2-\mathcal{F}({R}, {T})\right) T_{\mu \nu}-\mathcal{F}({R}, {T}) \Theta_{\mu \nu},
\label{5}
\end{align}
 with
 \begin{equation}
\mathcal{F}({R}, {T}) \equiv \frac{\partial f({R}, {T})}{\partial {T}} \quad \text { and } \quad F({R}, {T}) \equiv \frac{\partial f({R}, {T})}{\partial {R}},
\label{5'}
\end{equation}
where $\Box = \nabla^\mu\nabla_\mu$ is the D'Alembertian operator, $\nabla_\mu$ is the covariant derivative, and
\begin{equation}
\Theta_{\mu \nu} \equiv g^{\alpha \beta} \frac{\delta {T}_{\alpha \beta}}{\delta g^{\mu \nu}}=-2 {~T}_{\mu \nu}+g_{\mu \nu} {L}_{{m}}-2 g^{\alpha \beta} \frac{\partial^2 {~L}_{{m}}}{\partial g^{\alpha \beta} \partial g^{\mu \nu}}.
\label{teta}
\end{equation}

The field Eqs.  \eqref{5} reduce to the standard General Relativity when $f(R,T) \equiv R$.

To apply this modified gravity to cosmology, we consider the flat Friedmann-Lemaitre-Robertson-Walker (FLRW) line element given by
\begin{equation}
d s^2=-d t^2+a^2(t)\left(d r^2+r^2 d \theta^2+r^2  \sin\theta^2d\phi^2\right),
\label{4}
\end{equation}
where $a(t)$ is the scale factor and t represents the cosmic time. \\
In line with the existing literature, we choose $L_{\text{m}} = p$ \cite{Faraoni,Zahra,Akarsu}, and Eq. \eqref{teta} can be rewritten as:
\begin{equation}
\Theta_{\mu \nu} = -2 T_{\mu \nu} +p g_{\mu \nu}. 
\label{SAD}
\end{equation}
Additionally, the energy-momentum tensor, with its derivation provided in the Appendix \eqref{Appendix}, is expressed as that of a perfect fluid, i.e.,

\begin{equation}
T_{\mu\nu}=(\rho+p)u_{\mu}u_{\nu}+p g_{\mu\nu},
\label{mou}
\end{equation}

where $u_\mu$ is the four velocity.\\

Using Eqs. \eqref{SAD} and \eqref{mou}, we can rewrite Eqs. \eqref{5} as follows

\begin{align}
 &  3 H^2 F({R}, {T})+\frac{1}{2}(f({R}, {T})-F({R}, {T}) {R})+3 \dot{F}({R}, {T}) H   \nonumber  \\
  &\qquad =\left(\kappa^2+\mathcal{F}({R}, {T})\right) \rho+\mathcal{F}({R}, {T}) p,
  \label{9'}
\end{align}
 and
\begin{align}
& 2 F({R}, {T}) \dot{H}+\ddot{F}({R}, {T})-\dot{F}({R}, {T}) H\nonumber  \\
  &\qquad =-\left(\kappa^2+\mathcal{F}({R}, {T})\right)(\rho+p),
\label{10'}
\end{align}
 where $p$ and $\rho$ are the pressure and the energy density of the cosmic fluid, respectively.

\section{Running vacuum  model}
\label{Chap3}
To reach our objective, we consider a cosmological form of $f(R,T)$ as $f(R,T) =R+ 2 \lambda \kappa^2 T -2 \Lambda$, where $\lambda$ is a free parameter of the model and $\Lambda$ is the cosmological constant. The first and second Friedmann equations, Eqs. \eqref{9'} and \eqref{10'}, can be re-expressed as
\begin{equation}
\begin{aligned}
 3 H^2 +\frac{1}{2}( 2 \kappa^2 \lambda T-2\Lambda)  = \kappa^2\left(1+ 2\lambda \right) \rho+ 2 \kappa^2 \lambda p,
\end{aligned}
\end{equation}
and
\begin{equation}
2  \dot{H} = - \kappa^2 \left(1+2\lambda  \right)(\rho+p).
\end{equation}
In fact, Eq. \eqref{mou} provides the trace of the stress-energy tensor of matter i.e. $T=3p-\rho$ and the Friedmann equations take the following form
\begin{equation}
\left\{\begin{array}{c}
3 H^2=\kappa^2 \left(1+3 \lambda\right) \rho- \kappa^2 \lambda p + \Lambda, \\
-2 \dot{H}-3 H^2=\kappa^2 \left(1+3 \lambda\right) p-\kappa^2 \lambda \rho-\Lambda,
\end{array}\right.
\label{13}
\end{equation}
which can be expressed, by introducing an effective (geometrical DE) pressure $p_{\text{eff}}$ ($p_\Lambda$) and energy density $\rho_{\text{eff}}$ ($\rho_\Lambda$), as
\begin{equation}
\left\{\begin{array}{c}
3 H^2=\kappa^2 \rho_{\text{eff}}=\kappa^2\left(\rho+\rho_{\Lambda}\right), \\
-2 \dot{H}-3 H^2=\kappa^2 p_{\text{eff}}=\kappa^2\left(p+p_{\Lambda}\right),
\end{array}\right.
\label{35}
\end{equation}
where
\begin{equation}
\left\{\begin{array}{c}
\kappa^2\rho_{\text{eff}}=\kappa^2\left(\rho+ \lambda(3 \rho-p)\right)+\Lambda, \\
\kappa^2 p_{\text{eff}}=\kappa^2\left(p+ \lambda(3 p-\rho)\right)-\Lambda,
\end{array}\right.
\end{equation}
and 
\begin{equation}
\left\{\begin{array}{c}
\kappa^2 \rho_{\Lambda}=\kappa^2 \lambda(3 \rho-p)+\Lambda, \\
\kappa^2 p_{\Lambda}=\kappa^2 \lambda(3 p-\rho)-\Lambda.
\end{array}\right. \label{0è}
\end{equation}
Using Eqs. \eqref{13}  and  \eqref{0è}, $\rho_{\Lambda}$ and $p_{\Lambda}$ can be expressed as
\begin{equation}
\rho_{\Lambda}(H, \dot{H})=\frac{\lambda}{\kappa^2\left(1+4 \lambda\right)}\left(12 H^2+ \frac{2}{2 \lambda+1} \dot{H} + \frac{ \Lambda}{ \lambda}\right),
\label{rho_Lambda}
\end{equation}

\begin{equation}
p_{\Lambda}(H, \dot{H})= -\frac{\lambda}{\kappa^2\left(1+4 \lambda\right)}\left(12 H^2+2 \frac{8 \lambda +3}{2 \lambda+1} \dot{H}+\frac{ \Lambda}{ \lambda}\right).
\label{P_Lambda}
\end{equation}

We notice that the energy density, $\rho_{\Lambda}$, has the form of the RVE density, as provided in \cite{Sola2017}, and can be written as
\begin{equation}
\rho_{\Lambda}(H, \dot{H})=\frac{3}{\kappa^2}\left(  c_0 + \nu H^2+\frac{2}{3} \gamma \dot{H}\right),
\end{equation}
where $c_0=\frac{ \Lambda}{3\left(1+4 \lambda\right)}$, $\nu =  \frac{4 \lambda}{ \left(1+4 \lambda\right)}$, and $\gamma=\frac{\lambda}{ (2 \lambda+1) (1+4 \lambda)}$  are constants.
 The last two parameters characterize the RVE model and distinguish it from the $\Lambda$CDM model. Indeed, for $\lambda=0$, the form of $f(R,T)$ reduces to $R-2 \Lambda$, i.e. to $\Lambda$CDM and Eqs. \eqref{rho_Lambda} and \eqref{P_Lambda} describe the energy density and pressure of the vacuum i.e. $p_\Lambda=-\rho_\Lambda$.  These two parameters, which characterize the RVE model, are independent in the literature while in this equivalence they are derived from the $f(R,T)$ parameter, $\lambda$.
\subsection{Equation of state}
The equation of state of the running vacuum, from Eqs. \eqref{rho_Lambda} and \eqref{P_Lambda}, writes
\begin{equation}
p_{\Lambda}(H, \dot{H})= -\rho_{\Lambda}(H, \dot{H})- \frac{4\lambda}{\kappa^2 (1+2\lambda)}\dot{H}.
\label{EoSofRV}
\end{equation}
This expression deviates from the traditional equation of state of running vacuum energy, i.e. $p_{\Lambda}(H)=-\rho_{\Lambda}(H)$, by the term $ \frac{4\lambda}{\kappa^2 (1+2\lambda)}\dot{H}$. 
Recent studies, conducted by the authors of \cite{Moreno2022,Rezaei2022},  have shown that the equation of state of the running vacuum energy within the framework of quantum field theory is not exactly equal to $-1$. They have shown that the running vacuum energy becomes dynamical and evolves as a function of the Hubble parameter and its derivatives.
\\
By comparing Eq. \eqref{EoSofRV} with the results of references \cite{Moreno2022}, it can be inferred that the correction term, which depends only on $\dot{H}$, represents quantum corrections of order 2 to the standard equation of state which is valid only in the classical theory disregarding any influences from a quantum matter field.
We conclude that the energy density due to modified gravity represented by $f(R,T)=R+2 \kappa^2 \lambda T-2\Lambda$ takes the form of a running vacuum energy density with the following dynamical equation of state 
\begin{equation}
\omega_{\Lambda}(H, \dot{H})= -1- \frac{4\lambda}{\kappa^2 (1+2\lambda)}\frac{\dot{H}}{\rho_{\Lambda}(H, \dot{H})}.
\label{EoSofRV2}
\end{equation}
For $\lambda=0$, the parameters $\nu$ and $\alpha$, appearing in Eq. \eqref{rho_Lambda} and which are responsible for the running vacuum energy, vanish and we recover the general relativity with $\rho_\Lambda$ being the cosmological constant. 
The values of $\nu$ and $\alpha$ are naturally small in this context. 
So our modified gravity is equivalent to a  running vacuum energy in the context of general relativity.
\subsection{Energy-conservation}

In the literature, the interaction form between cosmic fluids and dark energy density is generally established from a phenomenological point of view due to the absence of a theory determining such a form. In our setup, the form of such interaction is inferred from the non conservation of the stress energy tensor in $f(R,T)$ gravity. Indeed, by replacing the form of $f(R,T)$ under study in Eq. \eqref{5} and by introducing an effective stress-energy tensor, $T^{\text{eff}}_{ \mu\nu}$, the gravitational field equation  can be rewritten in the form of Einstein's equations as  \cite{Moraes2018, Harko2011}
\begin{equation}
G_{ \mu\nu} = \kappa^2 T^{\text{eff}}_{ \mu\nu},
\label{18'}
\end{equation}
where \quad $T^{\text{eff}}_{ \mu\nu} = T_{\mu\nu} + \tilde{T}_{\mu\nu}$, with $ \tilde{T}_{\mu\nu}  \equiv \lambda [2(T_{\mu\nu} + pg_{\mu\nu} ) + T g_{\mu\nu}]$.\\

Nonetheless, the Bianchi identities, 
 $\nabla^{ \mu}G_{\mu\nu}=0$, gives
\begin{equation}
\dot{\rho}+3 H(\rho+p)=-\frac{\lambda}{\left(1+2 \lambda\right)}(\dot{\rho}-\dot{p}).
\label{7è}
\end{equation}

Furthermore, starting from Eqs. \eqref{0è}, we can observe that the running vacuum energy density, $\rho_{\Lambda}$,  is not conserved
\begin{equation}
\dot{\rho}_{\Lambda}+3 H(\rho_{\Lambda}+p_{\Lambda})=\frac{\lambda}{\left(1+2 \lambda\right)}(\dot{\rho}-\dot{p}).
\label{ooo}
\end{equation}

The term $Q=\frac{\lambda}{\left(1+2 \lambda\right)}(\dot{\rho}-\dot{p})$ represents the interaction form that measures the energy transfer between the energy density of the cosmic fluid components and the running vacuum energy density, as long as the effective energy density, $\rho_{\text{eff}}$, remains conserved  
\begin{equation}
    \dot{\rho}_{\text{eff}}+3 H(\rho_{\text{eff}}+p_{\text{eff}})= 0.\label{eff}
\end{equation}

We conclude that, by virtue of $f(R,T)$ gravity, the interaction form between the energy density of the cosmic fluid and the RVE density is well established as shown in Eqs. \eqref{7è} and  \eqref{ooo}.

To illustrate the behavior of such an interaction, we distinguish three cases of interaction between the running vacuum energy and the cosmic fluid.  We assume that the cosmic fluid consists of  CDM and radiation i.e. $\rho= \rho_m + \rho_r$ and $ p= p_m + p_r $.  The first case of interaction is between RVE and radiation, i.e. we consider that CDM remains conserved.  We label this case by model $A$. In the second case, labeled model $B$, RVE interacts with CDM while the radiation remains conserved. Finally, labeled by model $C$,  RVE interacts with both components of the cosmic fluid i.e. with CDM and radiation. 
To gather these three interactions, we write Eqs. \eqref{7è} and  \eqref{ooo} as follow 
\begin{equation}
\left\{\begin{array}{c}
\dot{\rho}_m+3 H(\rho_m+p_m)=-\frac{\lambda}{\left(1+2 \lambda\right)} \alpha (\dot{\rho}-\dot{p}),\\
\dot{\rho}_r+3 H(\rho_r+p_r)=-\frac{\lambda}{\left(1+2 \lambda\right)} (1- \alpha)(\dot{\rho}-\dot{p}),\\
\dot{\rho}_{\Lambda}+3 H(\rho_{\Lambda}+p_{\Lambda})=\frac{\lambda}{\left(1+2 \lambda\right)}(\dot{\rho}-\dot{p}),
\end{array}\right. \label{23'}
\end{equation}
where the constant $\alpha$ parameterize the three cases of interaction e.g.  $\alpha=0$, and $1$ represent the model $A$, and the model $B$, respectively. However an arbitrary value of $\alpha$  corresponds to model $C$. To solve this system of equations, we only need to solve the first two equations, taking into account that
$p_r={\rho_r}/{3}$ and $p_m=0$.  The expression of RVE is obtained from \eqref{0è}.

\section{Model Solutions}\label{Chap4}

 \subsection{Model $A$}

In this model, the cosmic fluid comprises both cold dark matter and radiation. We consider that the energy density of CDM remains conserved $(\alpha=0)$, while the radiation density is not conserved. The system of Eqs. \eqref{23'} provides the following solutions
\begin{equation}
\Omega_{m}(x) = \Omega_{m0} e^{-3 x},
\end{equation}
and
\begin{equation}
\Omega_{r}(x) = 3 \lambda  \Omega_{m0} e^{-3 x}+ \left(\Omega_{r0}-3 \lambda  \Omega_{m0}\right) e^{-\frac{12 (2 \lambda +1) x}{8 \lambda +3}},
\end{equation}
where  "$0$" refers to the current value, $x=\ln{(a)}$, and the dimensionless energy density of radiation and CDM are $\Omega_{r}=(8\pi G\rho_{r}/3H^2_0)$ and $\Omega_{m}=(8\pi G\rho_{m}/3H^2_0)$, respectively. \\

The Friedmann equation of the model $A$, from \eqref{13}, is given by
\begin{align}
    E^2(x)=&\;-\frac{(8 \lambda +3)}{3}  (3 \lambda \Omega_{m0}-\Omega_{r0}) e^{-12\frac{(2 \lambda +1) x}{8 \lambda +3}}
    \notag \\
    &\;+\left(8 \lambda ^2+6 \lambda +1\right) \Omega_{m0} e^{-3x}  + \Omega_{ \Lambda},
\end{align}
with the constrain at the current time, $\Omega_{\Lambda }=1-(3\lambda+1)\Omega_{m0} -((8\lambda+3)/3) \Omega_{r0}$.
\subsection{  Model $B$}

In this model, we make the assumption that radiation remains conserved $(\alpha=1)$, while the energy density of CDM is not conserved. The system of Eqs. \eqref{23'} yields the following solutions
\begin{equation}
\Omega_{m}(x) =\Omega_{m0} e^{-\frac{3 (2 \lambda +1) x}{3 \lambda +1}}+\frac{8 \lambda  \Omega_{r0}}{3 (6 \lambda +1)}  \left(e^{-\frac{3 (2 \lambda +1) x}{3 \lambda +1}}-e^{-4 x}\right),
\end{equation}
and
\begin{equation}
\Omega_{r}(x)= \Omega_{r0} e^{-4 x}.
\end{equation}

The Friedmann equation of the model $B$ can be expressed as follows
\begin{equation}
E^2(x)= (3 \lambda +1) \Omega_{m}(x) +\left(\frac{8 \lambda }{3}+1\right) \Omega_{r} (x)+ \Omega_{ \Lambda},
\end{equation}
with the constrain at the current time, $\Omega_{ \Lambda}=1-(3\lambda+1)\Omega_{m0} -((8\lambda+3)/3) \Omega_{r0}$.
\subsection{  Model $C$}

In this case, we explore the general situation where the budget of the cosmic fluid, i.e. cold dark matter and radiation,  engages  interactions with each others as in Eqs. \eqref{23'}.
\\

Through the resolution of Eqs. \eqref{23'}, we obtain the evolution of the energy density for both cold dark matter and radiation  as follows
\begin{equation}
\Omega_{m}(x) = \frac{I(x)+J(x)}{6 K} \exp{\left(-\frac{3 x\left(K+2 (\alpha +8) \lambda +7\right)}{2 (\alpha +8) \lambda +6} \right)}, \label{25}
\end{equation}
and
\begin{equation}
\Omega_{r}(x) =\frac{S(x)+R(x)}{2 K} \exp\left( {-\frac{3 x \left(K+2 (\alpha +8) \lambda +7\right)}{2 (\alpha +8) \lambda +6}} \right), \label{26}
\end{equation}
where
\begin{equation}
K=\sqrt{4 \alpha ^2 \lambda ^2+4 \alpha  (8 \lambda +3) \lambda +1},
\end{equation}
\begin{equation}
I(x)=16 \alpha  \lambda \Omega_{r0} \left(\exp \left(\frac{3 x K}{(\alpha +8) \lambda +3}\right)-1\right),
\end{equation}
\begin{align}
J(x)=&3 \Omega_{m0}\left(K+6 \alpha  \lambda +1\right) \exp \left(\frac{3 x K}{(\alpha +8) \lambda +3}\right) \nonumber  \\
&+3 \Omega_{m0}\left(K-6 \alpha  \lambda -1\right),
\end{align}
\begin{align}
S(x)=&3 \Omega_{r0}\left(K-6 \alpha  \lambda +1\right) \exp \left(\frac{3 x K}{(\alpha +8) \lambda +3}\right) \nonumber \\
&+3 \Omega_{r0}\left(K+6 \alpha  \lambda -1\right),
\end{align}
and
\begin{equation}
R(x)= 6 \lambda  (\alpha -1)\Omega_{m0} \left(\exp \left(\frac{3 x K}{(\alpha +8) \lambda +3}\right)-1\right).
\end{equation}

The Friedmann equation of model $C$ is written as
\begin{equation}
E^2(x)= (3 \lambda +1) \Omega_{m}(x) + (\frac{8 \lambda}{3} +1) \Omega_r(x)+\Omega_{ \Lambda}.
\end{equation}
with the constraint $\Omega_{ \Lambda }=1-(3\lambda+1)\Omega_{m0} -((8\lambda+3)/3) \Omega_{r0}$ at the current time.

\section{Data and methodology}\label{Chap5}
In order to test our three models, $A$,  $B$ and  $C$, we use the following observational data namely Pantheon+ \cite{Scolnic2022}, H(z) measurements \cite{Moresco2016}, cosmic microwave background (CMB) \cite{Aghanim2020} and baryon acoustic oscillations (BAO) \cite{Alam2017}. Furthermore, we perform  cosmological parameters extraction of these models, by running the Markov Chain Monte Carlo (MCMC) analysis \cite{Foreman2013},   and compare them with the $\Lambda$CDM model.\\

\subsection{Pantheon+ dataset}
We use the recent SNIa measurements from Pantheon$^+$ sample with 1701 light curves of 1550 SNIa in the redshift range of $z \in[0.001,  2.261]$  \cite{Scolnic2022}. We refer to this dataset as PP.\\
The chi-square of Pantheon+ dataset is given by
\begin{equation}
\chi_{\mathbf{PP}}^2= \overrightarrow{D}^T \cdot \mathbf{C}_{\text {PP}}^{-1} \cdot  \overrightarrow{D},
\label{"""}
\end{equation}
where $\mathbf{C}_{\text {PP }}$ is the covariance matrix of Pantheon+ data, and $\overrightarrow{D} = m_{{\text{Bi}}} - M - \mu_{{\text{th}}}$ \cite{Chaudhary2023}, with $m_{Bi}$  is the apparent  magnitude, $M$ is the absolute magnitude and $\mu_{{\text{th}}}$  the theoretical distance modulus.   Given a cosmological model, the theoretical distance modulus is given by
\begin{equation}
\mu_{{th}}(z)=5 \log _{10} \frac{D_L(z)}{\left(H_0 / c\right) {\text{Mpc}}}+25,
\end{equation}
where $H_0$ is the Hubble rate at the present time and c is the speed of light. The expression for $D_L(z)$ can be written as follows
\begin{equation}
D_L(z)=(1+z)  \int_0^z \frac{{d} z^{\prime}}{E\left(z^{\prime}\right)},
\end{equation}
where $E(z) = \frac{H(z)}{H_0}$ is the normalised Hubble rate.
\\

To break the degeneracy existing between Hubble constant, $H_0$,  and the absolute magnitude, $M,$ we consider SH0ES Cepheid measurements \cite{Riess2022}, which facilitate constraints on both parameters. In this case, the SNIa distance residuals can be modified  by redefining the vector $\overrightarrow{D}$ in Eq. \eqref{"""}, using the distance moduli of supernovae type Ia  in Cepheid hosts, as follow
\begin{equation}
\overrightarrow{D'_i} = \left\{
    \begin{array}{ll}
       m_{Bi} - M - \mu^{Ceph}_{i}& i \in \text{Cepheid \ hosts} \\
        m_{Bi} - M - \mu_{th}(z_i) & \mbox{ otherwise},
    \end{array}
\right.
\label{53}
\end{equation}
with $\mu^{Ceph}_{i}$ denotes the distance modulus associated with a Cepheid host. Therefore, Eq. \eqref{"""}  becomes
\begin{equation}
\chi_{\mathbf{PP}}^2= \overrightarrow{D'}^T \cdot \mathbf{C}_{\text {PP}}^{-1} \cdot  \overrightarrow{D'}.
\end{equation}
\subsection{$H(z)$  dataset}
We also use the Hubble parameter data consisting of 36 data  points \cite{Zhang2014, Jimenez2002} in the redshift range  $0.07 \leqslant z \leqslant 2.34$, where 30 data points are obtained using the differential age of galaxies method \cite{Joan2005} and 6 data points are extracted using the BAO radial size measurements and other methods \cite{Bautista2017}. The chi-square of these data, $\chi_{H(z)}^2$, is  defined as
\begin{equation}
\chi_{H(z)}^2=\sum_{i=1}^{36}\left[\frac{H_{o b s, i}-H\left(z_i\right)}{\sigma_{H, i}}\right]^2,
\end{equation}
where $H_{obs, i}$, $H\left(z_i\right)$ and $\sigma_{H, i}$  are  the observation value, the theoretical prediction, and the uncertainty of the Hubble parameter,   respectively.

\subsection{CMB data}
We use three cosmic microwave background (CMB) data points the baryon density, $\Omega_{\text{b}}h^2$,  the acoustic scale, $\ell_a$,  and the shift parameter, $R$. The theoretical expressions of the last two quantities are given respectively, by \cite{Eiichiro2009}
\begin{equation}
    \ell_a\equiv (1+z_{\text{cmb}})\frac{\pi D_A(z_{\text{cmb}})}{r_s(z_{\text{cmb}})},
\end{equation}
and
\begin{equation}
    R\equiv (1+z_{\text{cmb}})\sqrt{\Omega_m H_0^2}D_A(z_{\text{cmb}}),
\end{equation}
where $z_{\text{cmb}}$ is the redshift at the decoupling era, given by Planck data \cite{Aghanim2020} and depends weakly on $\Omega_b$ and $\Omega_m$ \cite{Hu1996}. The angular diameter distance, $D_A$, and the comoving sound horizon, $r_s$,  are given  respectively by
\begin{equation}
   D_A(z)=[H_0(1+z)]^{-1}\int^z_0\frac{dz'}{E(z')},\label{dA}
\end{equation}
\begin{equation}
   r_s(z)=H_0^{-1}\int^a_0 \frac{da'}{a'E(a')\sqrt{3(1+R_b)a'}},
\end{equation}
where,  $a=1/(1+z)$ and $R_b=31500\Omega_bh^2(T_{\text{cmb}}/2.7{\text{K}})$, with $T_ \text{cmb}=2.275{\text{K}}$ \cite{Fixsen2009}.
Finally, the contribution of CMB to $\chi_{{tot}}^2$ is as follows
\begin{equation}
\chi_{{{\text{cmb}}}}^2=P^{T}_{{{\text{cmb}}}}.\mathcal{C}^{-1}_{{{\text{cmb}}}}.P_{{{\text{cmb}}}},
\end{equation}
with $\mathcal{C}_{{{\text{cmb}}}}$ is the covariance matrix, given by \cite{Zhai2019}
\begin{equation}
\mathcal{C}_{{{\text{cmb}}}} = 10^{-8} \times
\begin{pmatrix}
1598.9554  & 17112.007 & -36.311179 \\
17112.007 & 811208.45 & - 494.79813\\
-36.311179 & -494.79813 & 2.1242182
\end{pmatrix},
\end{equation}
and  $P_{{{\text{cmb}}}}$ is the CMB parameters vector, based on Planck data \cite{Aghanim2020}, and derived by
\begin{equation}
P_{{{\text{cmb}}}} = 
\begin{pmatrix}
R - 1.74963 \\
\ell_a - 301.80845
\\
\Omega_{\text{b}}h^2-0.02237
\end{pmatrix}.
\end{equation}

\subsection{BAO data}
In addition to these data, we also use the Baryon Acoustic Oscillation (BAO) measurements. To use BAO data, it is necessary to involve the expression linking the Hubble parameter $H(z)$ and the angular diameter distance $D_A$, given by
\begin{equation}
D_v\equiv \left[(1+z)^2\frac{z}{H(z)}D_A^2(z)\right]^{1/3},
\end{equation}
with $D_v$ is the angle-average distance and $D_A$ is the angular diameter distance.\\
The redshift at the drag epoch, is given by \cite{Eisenstein1998} 
\begin{equation}
z_d\equiv \frac{1291(\Omega_mh^2)^{0.251}}{1+0.659(\Omega_mh^2)^{0.828}[1+A_1(\Omega_bh^2)^{A_2}]^{-1}},
\end{equation}
where
\begin{align}
 &\;A_1=0.313(\Omega_mh^2)^{-0.419}[1+0.607(\Omega_mh^2)^{0.674}], \quad \text{and}\notag \\  &\;A_2=0.238(\Omega_mh^2)^{0.223}.
\end{align}
The contribution of BAO to $\chi_{{tot}}^2$ is as follows
\begin{equation}
\chi_{{{\text{BAO}}}}^2=P^{T}_{{{\text{BAO}}}}.\mathcal{C}^{-1}_{{{\text{BAO}}}}.P_{{{\text{BAO}}}},
\end{equation}
where, $\mathcal{C}_{{{\text{BAO}}}}$ is the covariance matrix and $P_{{{\text{BAO}}}}$ is the difference vector between observational measurements (third column of Tab \eqref{jjjj}) and theoretical predictions. \\

In our analysis, we have used different BAO measurements at different redshifts, namely 6dFGS at $z=0.106$~\cite{Beutler2011}, SDSS DR7 MGS at $z=0.15$ \cite{Ross2015}, BOSS-LOWZ at $z=0.32$ \cite{Kazin2014}, BOSS-CMASS at $z=0.57$ \cite{Kazin2014}, WiggleZ at $z=0.44$, $0.60 $, $0.73$ \cite{Fixsen2009}, and BOSS-DR12 at $z=0.38$, $0.51$, $0.61$ \cite{Alam2017}. We should mention that BOSS-DR12 and WiggleZ are correlated, and their covariance matrices are given in \cite{Fixsen2009}.\\

\begin{table*}
\caption{Summary of the BAO data used in this work.}\label{jjjj}
\begin{center}
\begin{tabular}{|c|c|c|c|c|c|c|}
\hline   
\bf BAO name &\bf$z$  &\bf BAO expression &\bf BAO measurement &$\bf {\sigma_{BAO}}$  &$\bf r_{s}^{\textit{fid}}$  &$\bf Ref$ \\
\hline
\hline
6dFGS &$0.106$ &$\frac{r_s}{D_V}$  &$0.327$   &$0.015$  &$-$ &\cite{Beutler2011}\\       \hline                                       
SDSS DR7 MGS &$0.15$ &$D_V\frac{r_{s}^{fid}}{r_{s}}$ &$4.47$    &$0.16$    &$148.69$ &\cite{Ross2015} \\ [0.2cm] 
\hline 
BOSS-LOWZ &$0.32$  &$D_V\frac{r_{s}^{fid}}{r_{s}}$    &$8.47$    &$0.17$ &$149.28$  &\cite{Anderson2014} \\ [0.2cm]                                         
\hline
BOSS-CMASS &$0.57$  &$D_V\frac{r_{s}^{fid}}{r_{s}}$   &$13.77$    &$0.13$ &$149.28$  &\cite{Anderson2014} \\ [0.2cm]      
\hline               
WiggleZ &$0.44$  &$D_V\frac{r_{s}^{fid}}{r_{s}}$  &$1716$    &$83$ &$148.6$    & \cite{Kazin2014}  \\ [0.2cm] 
 &$0.60$  &   &$2221$ &$101$  & &\\ [0.2cm] 
 &$0.73$ &  &$2516$ &$86$   &   &\\ [0.2cm]   \hline  
 BOSS-DR12 &$0.38$&$D_A(1+z)\frac{r_{s}^{fid}}{r_{s}}$  &$1512.39$   &$25.00$  & $147.78$ & \cite{Alam2017} \\[0.2cm] 
 &  &$H\frac{r_{s}^{fid}}{r_{s}}$ &$81.2087$ &$2.3683$ & & \\[0.2cm]  \cline{2-5}
  &$0.51$  &$D_A(1+z)\frac{r_{s}^{fid}}{r_{s}}$ &$1975.22 $ &$30.10$          &&\\[0.3cm]
                             
 & &$H\frac{r_{s}^{fid}}{r_{s}}$ &$90.9029$ &$2.3288$ & & \\\cline{2-5}                   &$0.61$  &$D_A(1+z)\frac{r_{s}^{fid}}{r_{s}}$ &$2306.68 $  &$37.08$& &\\[0.3cm]
& &$H\frac{r_{s}^{fid}}{r_{s}}$ &$98.9647$ &$2.5019$ & & \\                              
\hline
\hline                           
\end{tabular}
\end{center}
\end{table*}

To estimate the parameters of our cosmological models, we define the total chi-square, $\chi^2_{tot}$, of the combined data as follows
\begin{equation}
\chi_{{tot}}^2 = \chi_{{PP}}^2 + \chi_{{H(z)}}^2 +\chi_{{CMB}}^2+\chi_{{BAO}}^2.
\end{equation}

To obtain the optimal constraints on the cosmological parameters, we employ the MCMC method \cite{Foreman2013} using CMB+\\BAO+PP+H data set. In our analysis, we take into account a 5-dimensional space, for Models $A$ and $B$, which includes \{M, $\Omega_{m0}$, $\Omega_{b}h^2$, $\lambda$, $H_{0}$\} with $\alpha=0$ and $1$, respectively. For the model C, we consider $\alpha$ as a free parameter. In this analysis, we have adopted the priors listed in the table \eqref{TBTT}
\begin{table}
\centering
{\caption{The priors imposed on the different cosmological parameters. }\label{TBTT}}
\begin{tabular}{c|c}
\hline
\multicolumn{1}{c|}{\bf Parameters} & \multicolumn{1}{c}{\bf Prior} \\
\hline      
$\Omega_{\textrm{m}}$ &[0.2, 0.5]
 \\[0.1cm]
 $\Omega_{\textrm{b}} h^{2}$ &[0.005, 0.1]
 \\[0.1cm]
 $H_0$   & [40, 100]
 \\[0.1cm]
 $M$ & [-20,-19] 
 \\[0.1cm]
$\lambda$ & [-1,1]
 \\[0.1cm]
$\alpha$ & [0,1]
 \\[0.1cm]
\hline
\end{tabular}
\end{table}

There are many comparison criteria for evaluating models and determining the best model among them, as well as ranking them based on their statistical ability to fit the observed data points \cite{Nesseris2010}. Among these criteria, we use the $\chi^2_{red}$, the Akaike Information Criterion (AIC) \cite{Akaike1974}, and the Bayesian Information Criterion (BIC) \cite{Schwarz1978}.
The $\chi^2_{red}$,  the  AIC  and the  BIC criteria are defined, respectively, as follows
\begin{align}
&\;\chi^2_{red}=\frac{\chi_{\min }^2}{N_d-N_p}, \quad AIC=\chi_{\min }^2+2 N_p, \qquad 
 \text{and} \notag \\
&\;{BIC}=\chi_{\min }^2+N_p \ln N_d,
\end{align}
where $\chi^2_{min}=-2 \ln(\mathcal{L}_{m})$, with $\mathcal{L}_{m}$ is the likelihood, $N_p$ is the number of free parameters and $N_d$ is the number of data. 

We also calculate two important quantities, $\Delta$AIC and $\Delta$BIC, defined respectively as follows
\begin{align}
&\; \Delta AIC=AIC_{\textrm{model}}-AIC_{\Lambda \textrm{CDM}}, \qquad \text{and}  \notag \\ &\;  \Delta BIC=BIC_{\textrm{model}}-BIC_{\Lambda\textrm{CDM}},
\end{align}
where we  consider $\Lambda$CDM as the reference model and a smaller value of AIC (BIC) indicates that the model is most supported by the observation data.
The selection rules of $\Delta$AIC and $\Delta$BIC stipulate that for 0$\leqslant\Delta$AIC$<$2, the model has nearly the same level of support from the dataset as the reference model. For 2$\leqslant\Delta$AIC$<$4, it indicates that the model with a high value of AIC has somewhat less support. While for $\Delta$AIC$\geq10$, it implies that the model has very little support. The same rules can be applied to BIC. 0$<\Delta$BIC$\leqslant$2  indicates insufficient evidence against the model with high value of BIC. For 2$<\Delta$BIC$\leqslant$6, it indicates that there is evidence against the model while for 6$\leqslant\Delta$BIC$<10$ indicates strong evidence against the model.

\section{Results and Discussion} \label{Chap6}
\begin{figure}
\centering
\includegraphics[scale=0.35]{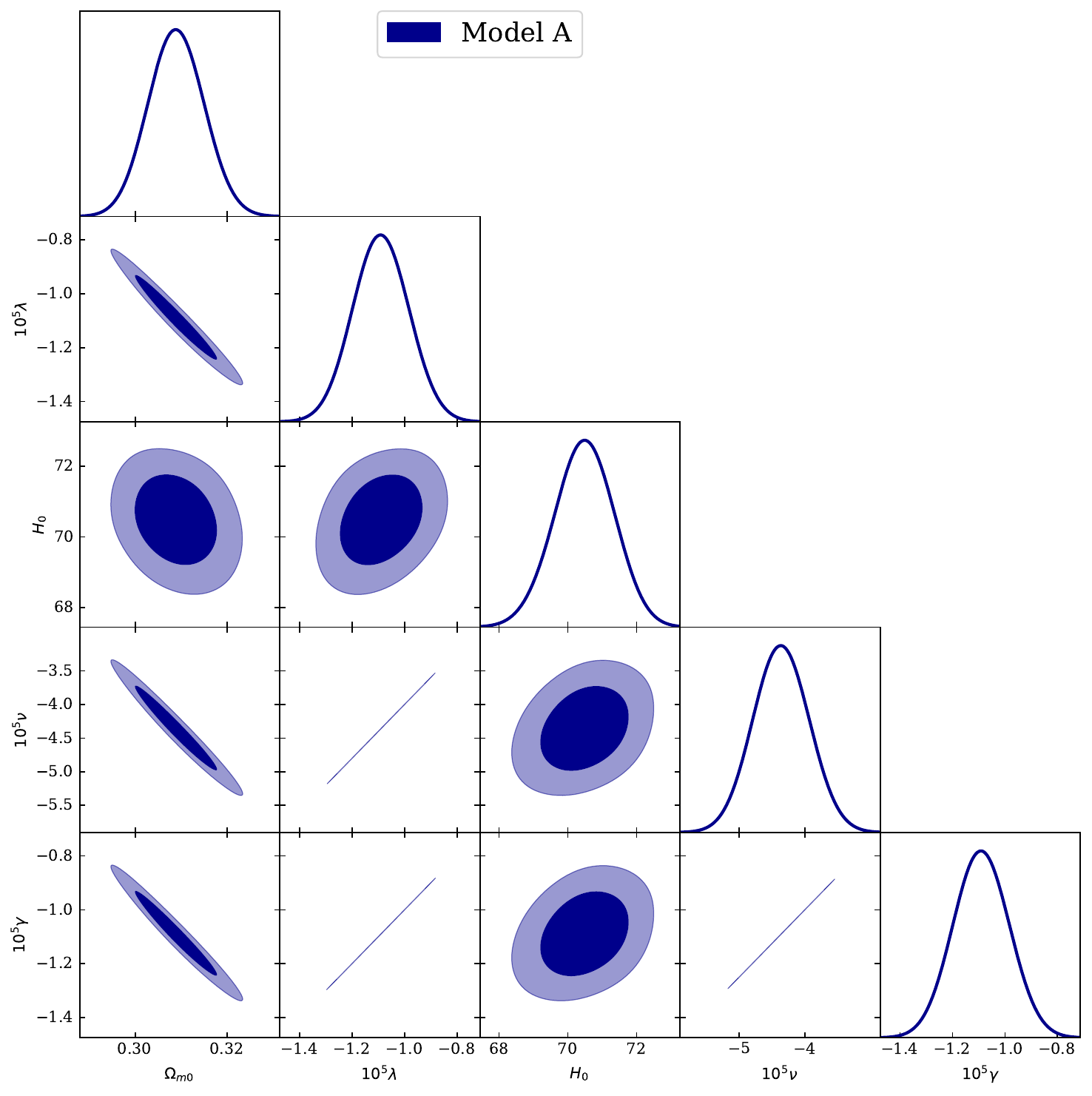}
\caption{The confidence contours at $1 \sigma$ and $2\sigma$ and  the 1D posterior distributions obtained from CMB+BAO+PP+H dataset for model A.}
\label{A}
\end{figure}

\begin{figure}
\centering
\includegraphics[scale=0.35]{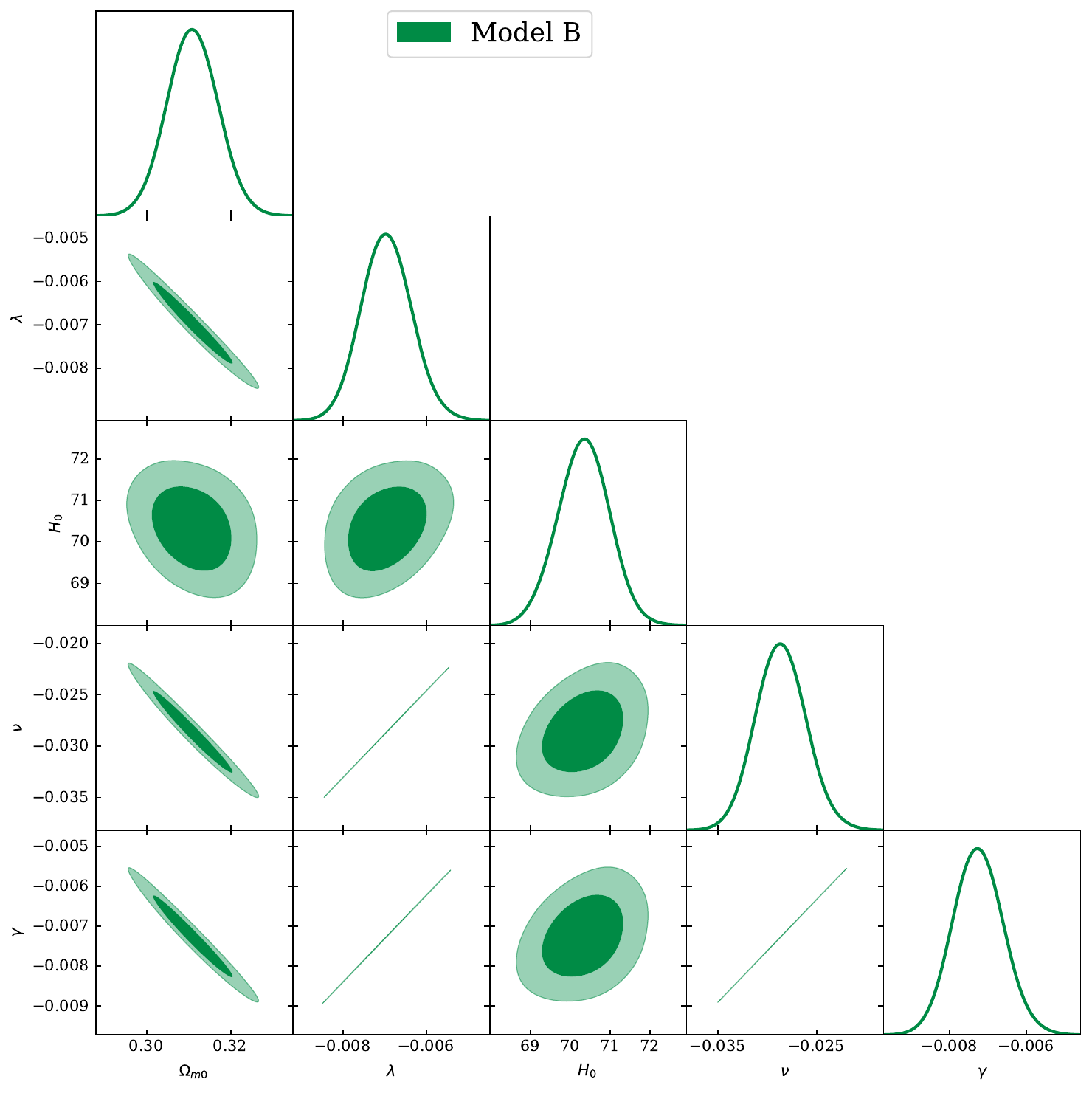}
\caption{The confidence contours at $1 \sigma$ and $2\sigma$ and  the 1D posterior distributions obtained from CMB+BAO+PP+H dataset for model B.}
\label{B}
\end{figure}

\begin{figure}
\centering
\includegraphics[scale=0.28]{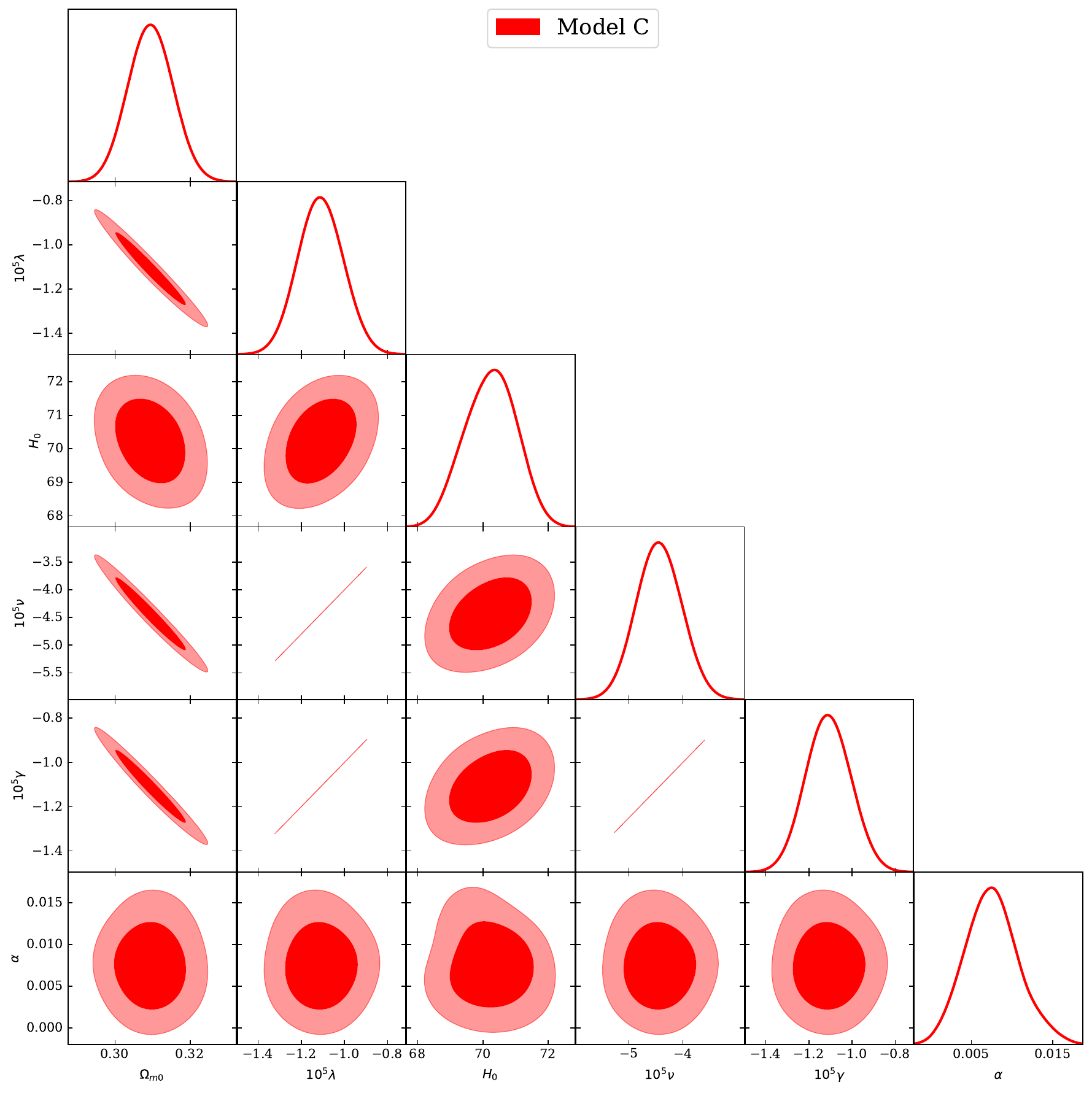}
\caption{The confidence contours at $1 \sigma$ and $2\sigma$ and  the 1D posterior distributions obtained from CMB+BAO+PP+H dataset for model C.}
\label{C}
\end{figure}

\begin{figure}
\centering
\includegraphics[scale=0.4]{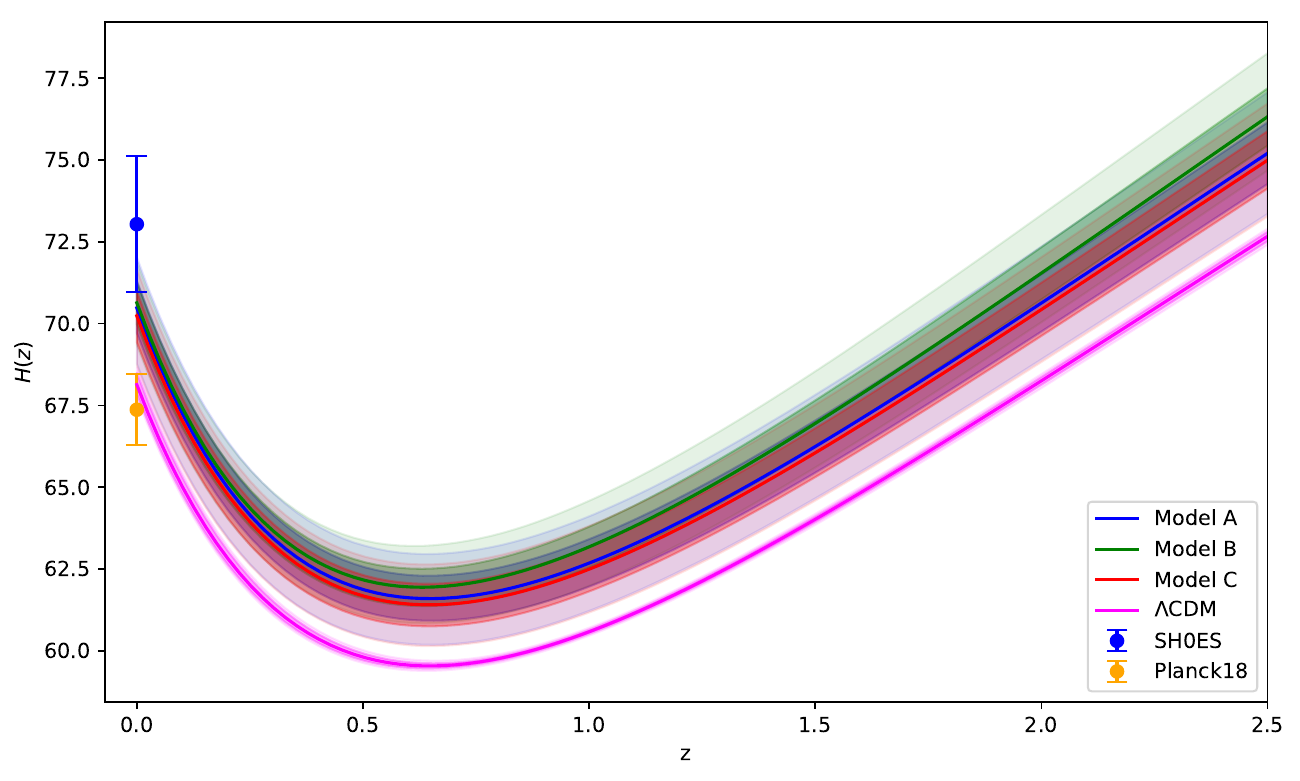}
\caption{Evolution of H(z) in a confidence interval $1\sigma$ and $2\sigma$ for A, B, C and $\Lambda$CDM models, using the chains of the free parameters obtained by CMB+BAO+PP+H dataset. Additionally, the SH0ES \cite{Riess2022} and Planck18 \cite{Aghanim2020} data points are shown at $2\sigma$.}
\label{G}
\end{figure}

 In Table \eqref{P},  we show the mean$\pm1\sigma$ of the cosmological parameters for $A$, $B$, $C$ and $\Lambda$CDM models, using CMB+BAO+\\PP+H dataset, as well as the $\chi^2_{tot}$, $\chi^2_{red}$,  $\Delta$AIC and $\Delta$BIC corresponding to each model. Figs. \eqref {A}, \eqref{B},  and \eqref{C} show the $2D$ marginalized confidence contours and $1D$ posterior distributions  for models $A$, $B$ and $C$, respectively.
\begin{table*}
\caption{
Summary of the mean values at $1\sigma$ of the free cosmological parameters, derived parameters, $\chi^2_{\text{min}}$ , $\chi^2_{\text{red}}$ as well as $\Delta$AIC and $\Delta$BIC  for each model.}
\centering
\begin{tabular}{c|c|c|c|c}
\hline
\hline
\multicolumn{1}{c|}{Data} & \multicolumn{4}{c}{CMB+BAO+PP+H }\\
\hline
\multicolumn{1}{c|}{Model} & \multicolumn{1}{c|}{$\Lambda$CDM} & \multicolumn{1}{c|}{Model A}&\multicolumn{1}{c|}{Model B}&\multicolumn{1}{c}{Model C}\\
\hline    
 
$\Omega_{\text{m0}}$   &  $0.3096^{+0.0041}_{-0.0027}$   & $0.3089\pm 0.0057$ &  $0.3109\pm 0.0060$  & $ 0.3094\pm 0.0059$   \\[0.1cm]
$\Omega_{\text{b}}h^2$   &  $0.02248\pm 0.00012$   & $0.02245\pm 0.00014$ &  $0.02284\pm 0.00012$  & $0.02245\pm 0.00014$   \\[0.1cm]
H$_0$[km s$^{-1}$ Mpc$^{-1}$]   &  $68.09^{+0.20}_{-0.30}$   & $ 70.48\pm 0.82$ &  $70.34\pm 0.64$  & $ 70.24\pm 0.78$   \\[0.1cm]
$\lambda$  & $-$   & $\left(\,-1.09\pm 0.10\,\right)\cdot 10^{-5}$ &  $-0.00696\pm 0.00059$  & $\left(\,-1.11\pm 0.10\,\right)\cdot 10^{-5}$   \\[0.1cm]
$\alpha$  &  $-$   & $0$ &  $1$  & $0.0075^{+0.0026}_{-0.0032}$   \\[0.1cm]
M$_{\text{B}}$ [mag]  &  $-19.4172^{+0.0065}_{-0.0083}$   & $-19.343\pm 0.025$ &  $-19.348\pm 0.019$  & $-19.351^{+0.026}_{-0.023}$   \\[0.1cm]
\hline 
\multicolumn{5}{c}{derived parameters}\\
\hline
$\nu$  &  $-$   & $\left(\,-4.36\pm 0.40\,\right)\cdot 10^{-5}$ &  $-0.0286\pm 0.0025$  & $\left(\,-4.44\pm 0.41\,\right)\cdot 10^{-5}$   \\[0.1cm]
$\gamma$&   $-$   & $\left(\,-1.09\pm 0.10
\,\right)\cdot 10^{-5}$ &  $-0.00726\pm 0.00064$  & $\left(\,-1.11\pm 0.10\,\right)\cdot 10^{-5}$   \\[0.1cm]
\hline 
\multicolumn{5}{c}{statistical results}\\
\hline
$\chi^2_{\text{min}}$  & $1593.89$  & $1584.71$ &  $1589.85$  & $1583.59$   \\[0.1cm]
$\chi^2_{\text{red}}$  & $0.913$  &  $0.908$  &  $0.911$  &  $0.908$   \\[0.1cm]

$\Delta$AIC  & $0$  & $-7.1$ & $-2.04$ & $-6.3$  \\[0.1cm]
$\Delta$BIC & $0$  & $-1.7$ & $+3.43$ &$+4.64$  \\[0.1cm]
\hline
\hline
\end{tabular}

\label{P}
\end{table*}
 From table \eqref{P},  we get $H_0= 68.09^{+0.20}_{-0.30}$ km s$^{-1}$ Mpc$^{-1}$ for $\Lambda$CDM and it is noticeable that nearly all models, A, B and C almost share the same current value of the Hubble parameter, which is approximately $H_0\simeq 70$ km s$^{-1}$ Mpc$^{-1}$ using CMB+BAO+PP+\\H dataset. Taking into account the uncertainty of  $H_0$, the difference between the values obtained for our models and the one obtained by $\Lambda$CDM is at $\simeq  2.7\sigma$. On the other hand, the interaction of the running vacuum increases the value of $H_0$ by $\simeq 3.5\%$, which may be a promising proposition for the Hubble tension problem (see Fig \eqref{G}).

 Moreover, for Model A (i.e. $\alpha=0$), we obtain a low value of $\lvert\lambda\rvert$ around $10^{-5}$, which justifies the very low interaction between radiation and the running vacuum. The same result is obtained for $\lambda$  when we consider $\alpha$ as a free parameter (i.e. model C). In addition to the low value obtained for $\lambda$, we also obtain a low value for $\alpha=0.0075^{+0.0026}_{-0.0032}$, which differs by $2.3\sigma$ from $\alpha=0$ (i.e. model A). These results show that the dataset used in our analysis prefers weak interaction between the running vacuum energy and the other components. We also notice that for all models, we obtain a negative value of  $\lambda$. At the present time, the interaction term $Q$ between RVE and matter, radiation or both of them, Eq. \eqref{ooo}, can be expressed as follows
 \begin{equation}
     Q =-\frac{\left(3 H_0^3 \lambda \right) \left(9 \Omega _{m_0}+8 \Omega _{r_0}\right)}{k^2 (\alpha  \lambda +8 \lambda +3)}.
 \end{equation}
  Since $\lambda$ is negative for all models, the interaction term is always positive, which means that the energy density is transferred from matter, radiation or both of them to RVE for model $A$, $B$ and $C$, respectively.  In table \eqref{P}, we also list the two parameters characterizing the running vacuum model, $\nu$ and $\gamma$, as derived parameters from $\lambda$. As expected, we have obtained a very small value of $\lvert\nu\rvert\ll 1$ and $\lvert\gamma\rvert\ll 1$, especially for models A and C, where we obtained $\lvert\nu\rvert\ll 10^{-4}$ and $\lvert\gamma\rvert\ll 10^{-4}$ as it is found in the literature \cite{Sarath2022}. As a result, models A and C remain very close to $\Lambda$CDM  thanks to the slight interaction between the running vacuum and the other components. \\

\begin{figure}
\centering
\includegraphics[scale=0.3]{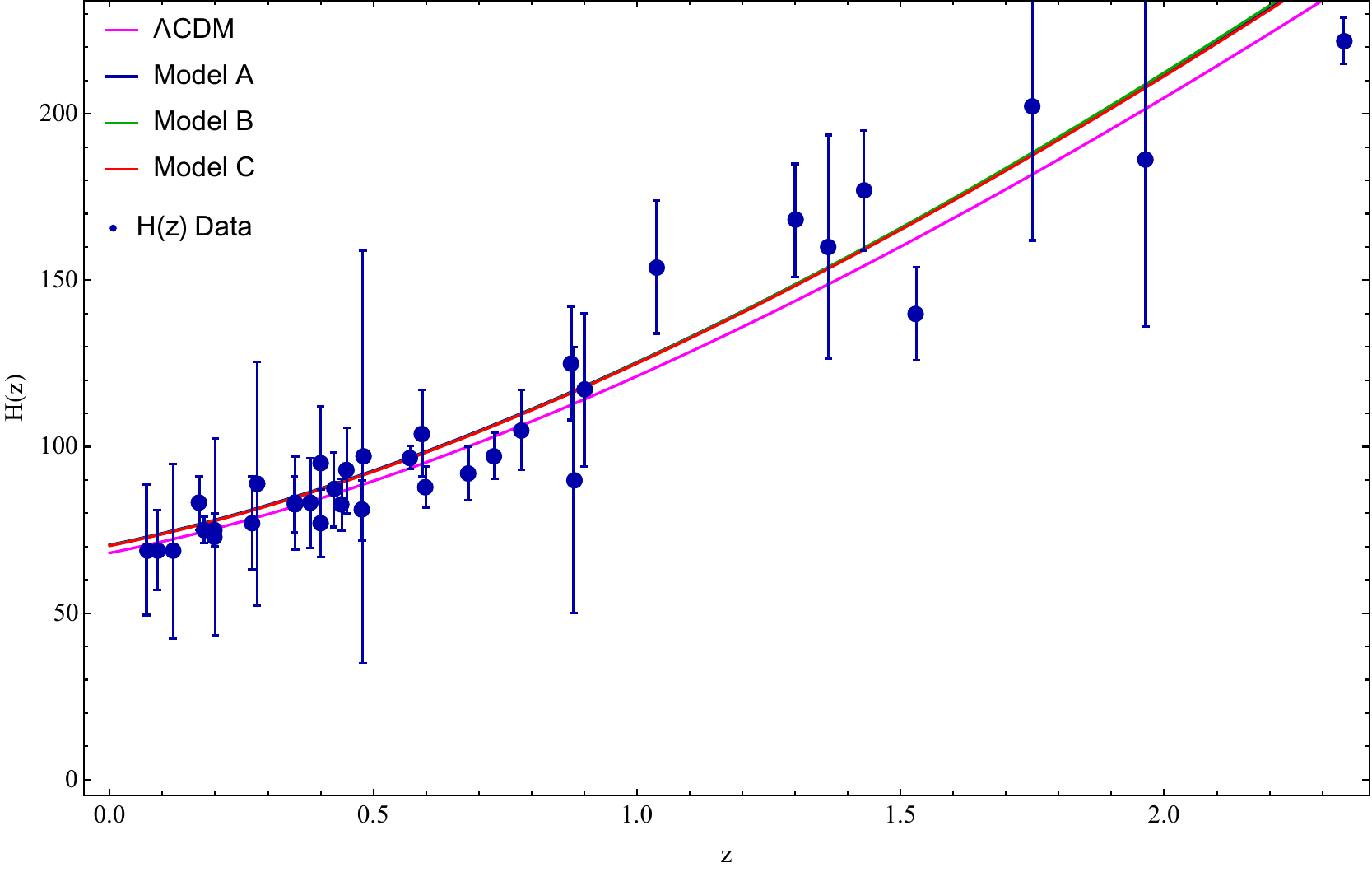}
\includegraphics[scale=0.3]{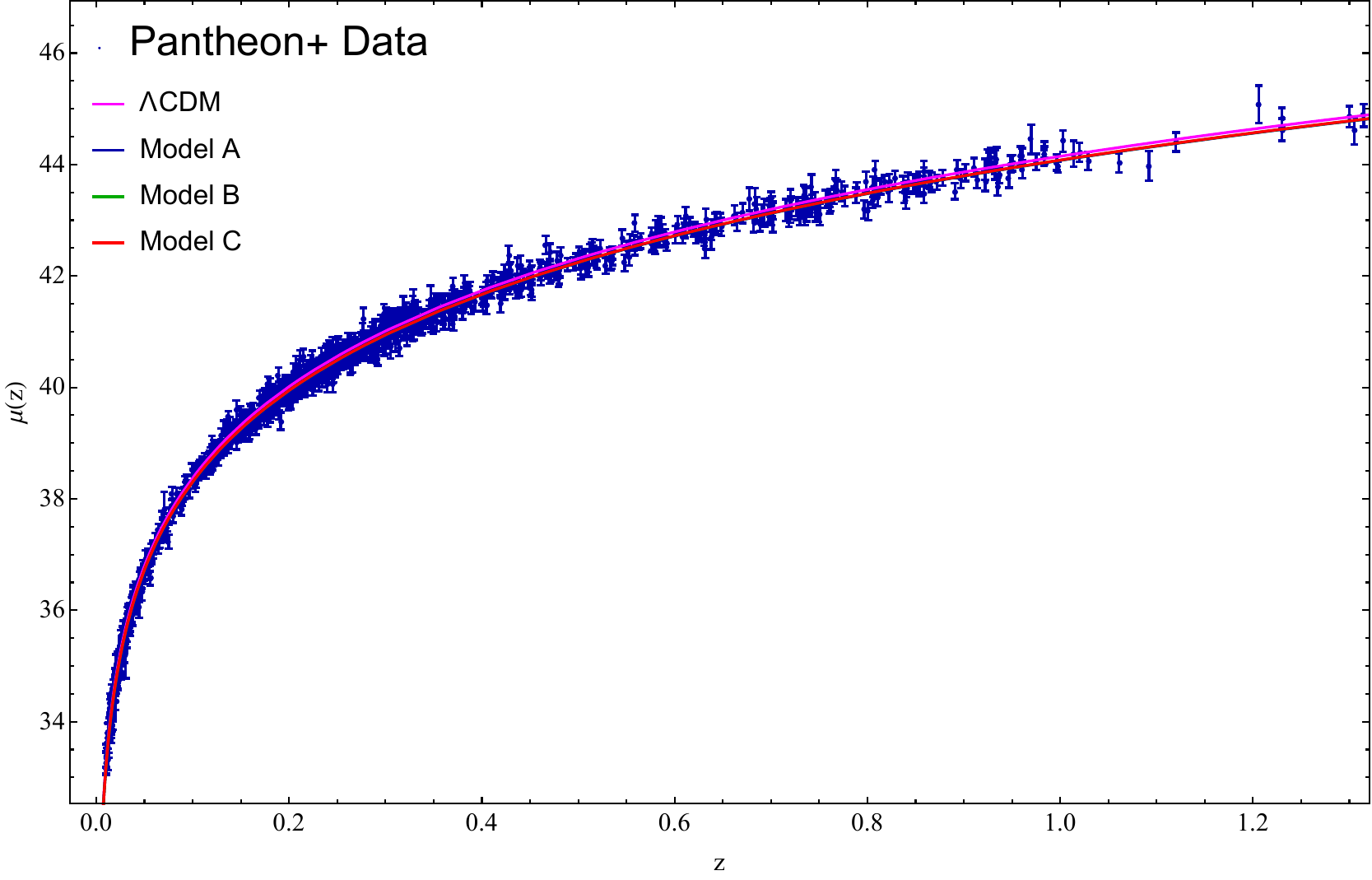}
\caption{Evolution of  H(z) and  the distance modulus $\mu(z)$ of the model $A$, $B$, $C$ and  $\Lambda$CDM  as a function of the redshift z.}
\label{3''}
\end{figure}

After estimating the best fit of model's parameters, we compare our models and $\Lambda$CDM with respect to Hubble’s 36 data points and Pantheon+ datasets by plotting the evolution of the Hubble parameter $H(z)$ and the distance modulus $\mu(z)$, respectively. This is shown in Fig. \eqref{3''} where we observe that all models are compatible with both datasets.\\

Table \eqref{P} shows also the values of $\chi^2_{\text{min}}$,  $\chi^2_{\text{red}}$, $\Delta$AIC and $\Delta$BIC. For models A and C, we get a negative value for $\Delta$AIC and  $\lvert\Delta \text{AIC}\rvert>6$. This means that models $A$ and $C$ have minimum AIC values and strong evidence suggesting that these models are best suited to the dataset used in our analysis compared to the $\Lambda$CDM model using the AIC criteria. While for model $B$ the difference between $\text{AIC (model $B$)}$ et $\text{AIC}$ $(\Lambda\text{CDM})$ is not enough to select the preferred model. The BIC criterion reduces the difference between $\Lambda$CDM and model $A$ and gives a positive value of $\Delta$BIC for models $B$ and $C$, which means that this criterion always favors models with a minimum number of parameters.\\

    The expression of the deceleration parameter, $q(z)$, which is a measure of the cosmic acceleration of the expansion of the Universe, is defined in terms of redshift as follows \cite{Tiwari2018}
\begin{equation}
q(z) =-\left(1- (z+1) \frac{ H'(z)}{H(z)}\right),
\end{equation}
where the prime denotes the derivative with respect to redshift $z$.\\


The equation of state (EoS) parameter represents the ratio of pressure to energy density i.e. it determines the relation between energy density and pressure in a particular phase of the Universe.  The expression of the effective EoS parameter, using $\omega_{\text{eff}}=p_{\text{eff}}/\rho_{\text{eff}}$, Eq. \eqref{35} and Eq. \eqref{eff},  is given by
\begin{equation}
\omega_{\text{eff}}(z)= -1+ \frac{2}{3} \frac{ H'(z)}{H(z)} (z+1).
\end{equation}

\begin{figure}
\centering
\includegraphics[scale=0.3]{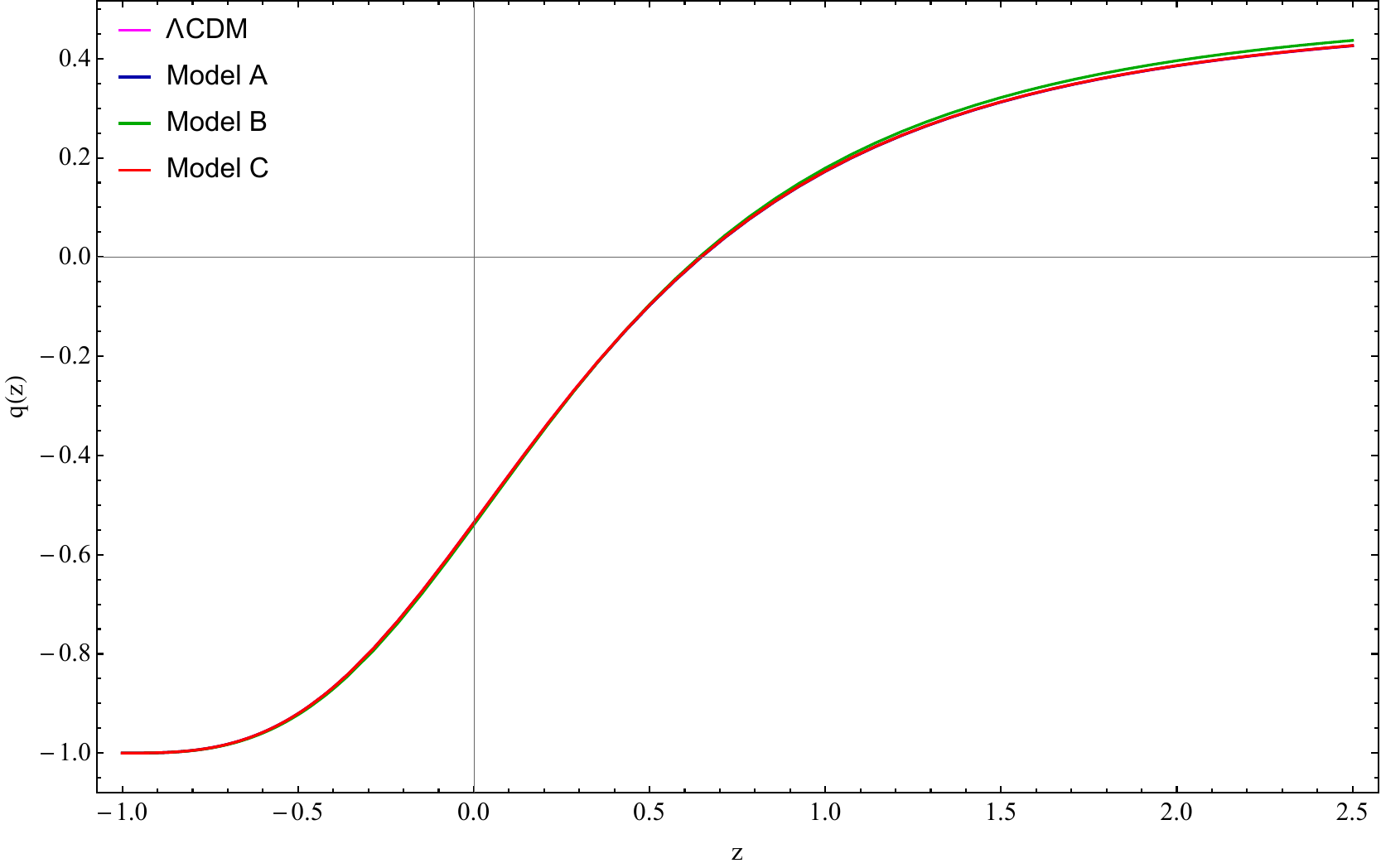}
\includegraphics[scale=0.3]{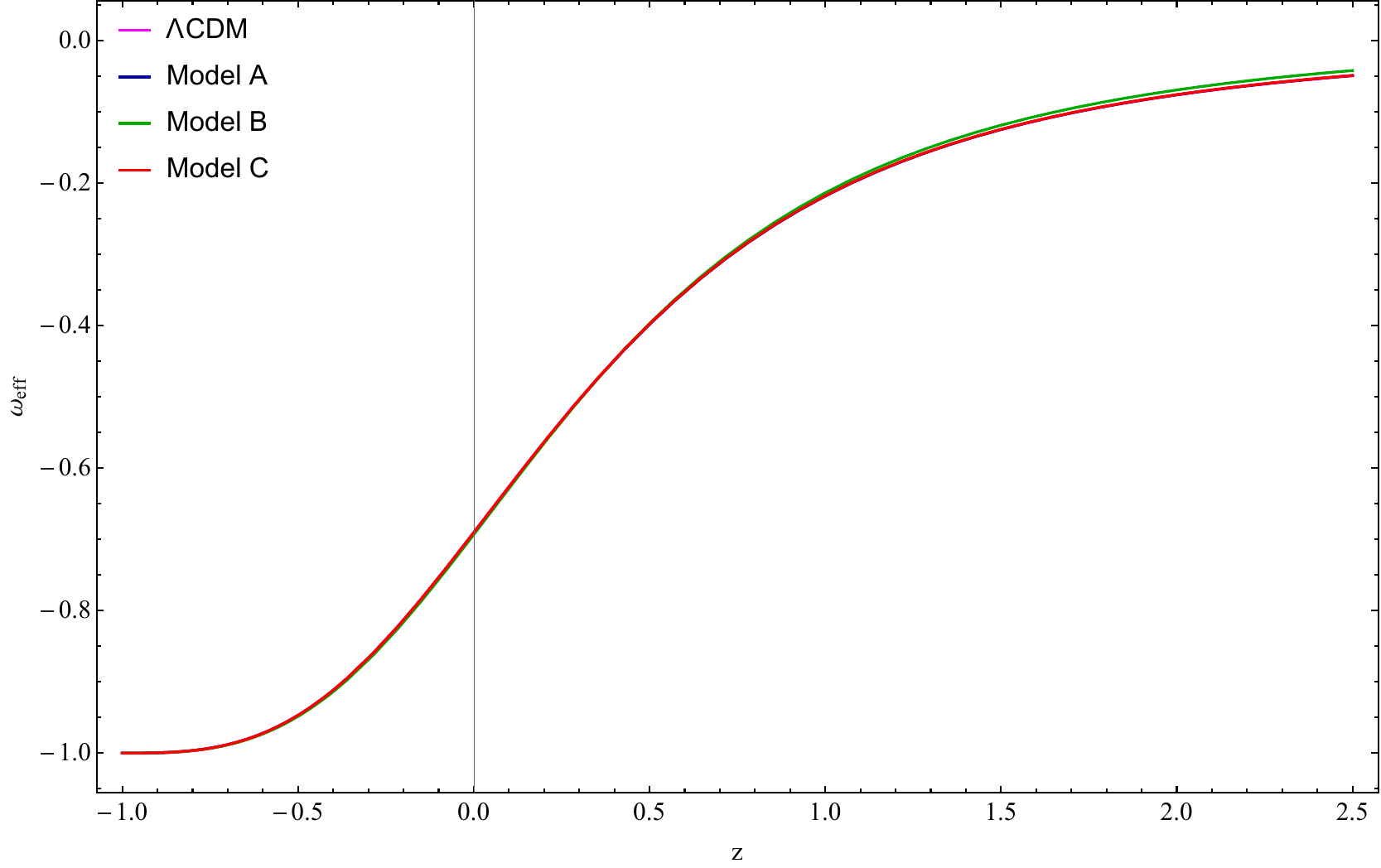}
\caption{Evolution of deceleration parameter $q(z)$ and of the equation of state parameter $\omega_{\text{eff}}$ of models  $A$, $B$, $C$ and $\Lambda$CDM as a function of the redshift.}
\label{8}
\end{figure}
\begin{table*}
\caption{Summary of the mean$\pm 1\sigma$ values of the $q_{\text{0}}$, $\omega_{{\text{eff,0}}}$ and $z_{\text{tr}}$ .}
\centering
\begin{tabular}{c|c|c|c|c}
\hline
\hline
Parameters & $\Lambda$CDM & Model $A$ & Model $B$& Model $C$\\ 
\hline
$q_{\text{0}}$           & $-0.5366\pm 0.0054$ & $-0.5364\pm 0.0085$ & $-0.540\pm 0.008$ & $-0.5357\pm 0.0087$ \\ 
$\omega_{{\text{eff,0}}}$           & $-0.691\pm 0.003$ & $-0.6909\pm 0.0057$ & $-0.6933\pm 0.0054$& $-0.6905\pm 0.00583$ \\ 
$z_{\text{tr}}$           & $0.6478\pm 0.0094$ & $0.6473\pm 0.0146$ & $0.6417\pm 0.0148$ & $0.6461\pm 0.0149$ \\ 
\hline
\hline
\end{tabular}

\label{JJ}
\end{table*}
\vspace{1.mm}
 Fig \eqref{8} illustrates the behavior of the deceleration parameter and  $\omega_{{\text{eff,0}}}$ of  the model $A$, $B$, $C$  and  $\Lambda$CDM.  In the redshift range  $ z \in [-1,2.5]$, the behavior of the deceleration parameter and the effective EoS parameter for all models closely resemble to that of the $\Lambda$CDM  model. In the Tab. \eqref{JJ}, we present the mean$\pm 1\sigma$ of $q_{\text{0}}$, $\omega_{{\text{eff,0}}}$ and $z_{\text{tr}}$ for all models. We obtain the same mean$\pm 1\sigma$ of the present value of deceleration $q_{\text{0}}$ and the present value of the effective EoS parameter. We see that the Universe underwent a transition phase from a deceleration to an acceleration expansion, where all models share the same value of $z_{\text{tr}}$  with a difference of $0.03\sigma$,  $0.3\sigma$ and  $0.09\sigma$ for models $A$, $B$ and $C$ compared to $\Lambda$CDM.
\section{Conclusions}
\label{Chap8}
In this paper, we introduced an analysis of $f(R,T)$ gravity formalism, focusing on an exploration of the energy-momentum tensor $T_{\mu \nu}$. Our findings indicate that the model described by $f(R,T)=R+2\kappa^2 \lambda T-2\Lambda$ is equivalent to the standard model incorporating the concept of running vacuum energy (RVE). Describing an effective dynamical dark energy, RVE engages an interaction with various components of the cosmic fluid, such as dark matter and radiation. In model $A$, we maintain the conservation of matter energy density while the energy density of radiation is not conserved. In model $B$, the energy density of matter is not conserved, while radiation energy density remains conserved. Finally, Model $C$ signifies scenarios where the energy densities of matter and radiation are not conserved.  To estimate the cosmological parameters of models $A$, $B$ and $C$, we performed a Markov Chain Monte Carlo analysis using the dataset combination CMB, BAO, $H(z)$ and  Pantheon+. By using the best fit values of the model parameters, we have analyzed the behavior of different cosmological parameters as the deceleration, and the effective equation of state. From Fig. \eqref{8}, we observed that the behavior of the deceleration parameter and the effective EoS parameter for all models is close to that of the $\Lambda$CDM  model. We have also demonstrated that the interaction term, $Q$ is positive at the present time, indicating that energy density is being transferred from matter, radiation, or both of them to RVE. Furthermore, the two parameters characterizing the running vacuum model are very small as expected. The results validate the credibility of the proposed cosmological models i.e. the validity of the equivalence between $f(R,T)$ gravity under consideration and the interacting running vacuum energy density in context of general relativity. As RVE is sourced from quantum field theory,  this equivalence may indicate a relation between $f(R, T)$ gravity and the quantum field theory.  Finally, we find that the equivalence between interacting running vacuum and $f(R,T)$ gravity increases the value of current Hubble rate, $H_0$ by  $3.5\%$, which may be a promising study, in the future, of the Hubble tension issue. 
\section*{ACKNOWLEDGMENTS}
We would like to thank T. Harko for his valuable comments and discussions on the results of this work.
\section*{APPENDIX
} \label{Appendix}
Consider a fluid described locally in a comoving inertial frame by various thermodynamical variables: internal energy, $U$, temperature, $T$, entropy, $S$, pressure, $P$, volume, $V$, chemical potential, $\mu$, and number of particles, $N$. We assume that there are no creation or annihilation processes, so that the particle number is conserved. \\
The first law of thermodynamics is written as \cite{Brown, Zahra}
\begin{equation}
    dU = T dS - P dV + \mu dN, \label{first}
\end{equation}
and the Gibbs-Duhem equation is
\begin{equation}
U = TS - PV + \mu N. \label{Gibbs}
\end{equation}
By defining  the particle number density, $n=N/V$, entropy per particle, $s=S/N$, and the energy density, $\rho=U/V$, the first law, Eq. (\ref{first}), and the Gibbs-Duhem relation, Eq. (\ref{Gibbs}), simplify, respectively as follows
\begin{equation}
d\rho = T n ds + \mu' dn, \label{drho}
\end{equation}
and
\begin{equation}
\rho = \mu' n - P, \label{rho}
\end{equation}
where $\mu' = \mu + T s$. 

Using the differential of the Gibbs-Duhem relation, and the first law of thermodynamics, we derive
\begin{equation}
dP = s n dT + n d\mu = n d\mu' - n T ds, \label{dp}
\end{equation}
which implies that $\rho = \rho(s, n)$ and $P = P(\mu', s)$.

We then define the particle number flux as \cite{Brown}
\begin{equation}
J^\mu = \sqrt{-g} n u^\mu, \label{Jmu}
\end{equation}
and the Taub current as \cite{Brown}
\begin{equation}
V^\mu = \mu' u^\mu, \label{Vmu}
\end{equation}
where $u^\mu$ is the fluid 4-velocity, and the particle number density, $n$, is related to the particle number flux by
\begin{equation}
n = \sqrt{\frac{g_{\mu\nu}J^\mu J^\nu}{g}}.\label{n}
\end{equation}

  With the above definitions, we obtain
\begin{equation}
J \equiv \sqrt{-J_\mu J^\mu} = \sqrt{-g} n, \quad\text{with}\quad J^\mu = J u^\mu, \label{J}
\end{equation}
and
\begin{equation}
V \equiv \sqrt{-V_\mu V^\mu} = \mu', \quad\text{with}\quad V^\mu = V u^\mu. \label{V}
\end{equation}
The entropy and particle production rates remain unchanged during the dynamical evolution. Therfore the variations of the entropy density, $s$, and the ordinary matter number flux vector density,
$J_\mu$, satisfy $\delta s=0$ and $\delta J_\mu=0$. It should also be noted that the Taub current $V_\mu$ is conserved i.e. $\delta V_\mu=0$ \cite{Brown}.

Applying the previously results, we vary the particle number density, $n$, and we find
\begin{align}
\delta n =&\frac{n}{2} (-g) u^\mu u^\nu \left( \frac{\delta g_{\mu\nu}}{g} - \frac{g_{\mu\nu}}{g^2} \delta g \right) \nonumber  \\
&= \frac{n}{2} \left( u_\mu u_\nu + g_{\mu\nu} \right) \delta g^{\mu\nu} \label{delta n}.
\end{align}
To derive the variations of energy density and pressure with respect to the metric, it is essential to calculate the variation $\delta\rho$ and $\delta P$.\\
In the context of isentropic processes, $\delta\rho$ and $\delta P$ are expressed as follows
\begin{equation}
\delta \rho = \frac{\rho + P}{n} \delta n, \label{delta rho}
\end{equation}
and
\begin{equation}
\delta P = n \, \delta\mu'.\label{delat p}
\end{equation}
The variation of $n$ is given by Eq. (\ref{delta n}), while from  Eq. (\ref{delat p}) the variation of $\mu'$  can be obtained as
\begin{equation}
\delta \mu' = \delta V = -\frac{V_\mu V_\nu}{2V} \delta g^{\mu\nu} 
= -\frac{1}{2} \mu' u_\mu u_\nu \delta g^{\mu\nu}.
\end{equation}
These relations provide the thermodynamic variations of the energy density and pressure with respect to the metric as
\begin{equation}
\frac{\delta \rho}{\delta g^{\mu\nu}} = \frac{1}{2} (\rho + P)(g_{\mu\nu} + u_\mu u_\nu), \label{delta rhog}
\end{equation}
and
\begin{equation}
\frac{\delta P}{\delta g^{\mu\nu}} = -\frac{1}{2} (\rho + P) u_\mu u_\nu,\label{delat pg}
\end{equation}
respectively. 

From the definition of the energy-momentum tensor  

\begin{equation}
T_{\mu\nu} = -\frac{2}{\sqrt{-g}} \frac{\delta (\sqrt{-g} L_m)}{\delta g_{\mu\nu}},
\end{equation}
 we have 
\begin{equation}
T_{\mu\nu} = L_m g_{\mu\nu} - 2 \frac{\delta L_m}{\delta g_{\mu\nu}}.\label{Tmunu}
\end{equation}
Now, by taking $L_m = P$, in the above equation and using Eq. (\ref{delat pg}), we find the energy momentum tensor for perfect fluid 
\begin{equation}
T_{\mu\nu}= (\rho + P) u_\mu u_\nu+P g_{\mu\nu}.
\end{equation}
The same result can be derived by setting $L_m=-\rho$ in Eq. (\ref{Tmunu}) and applying Eq. (\ref{delta rhog})


\end{document}